# Discovery of a quantum limit Chern magnet TbMn$_6$Sn$_6$

**Authors:** Jia-Xin Yin[1]*†, Wenlong Ma[2]*, Tyler A. Cochran[1]*, Xitong Xu[2]*, Songtian S. Zhang[1], Hung-Ju Tien[3], Nana Shumiya[1], Guangming Cheng[4], Kun Jiang[5], Biao Lian[6], Zhida Song[7], Guoqing Chang[1], Ilya Belopolski[1], Daniel Multer[1], Maksim Litskevich[1], Zijia Cheng[1], Xian Yang[1], Bianca Swidler[1], Huibin Zhou[2], Hsin Lin[8], Titus Neupert[9], Ziqiang Wang[5], Nan Yao[4], Tay-Rong Chang[3,10,11], Shuang Jia[1,12,13]†, M. Zahid Hasan[1,14]†

**Affiliations:**

[1]Laboratory for Topological Quantum Matter and Advanced Spectroscopy (B7), Department of Physics, Princeton University, Princeton, New Jersey, USA.

[2]International Center for Quantum Materials and School of Physics, Peking University, Beijing, China.

[3]Department of Physics, National Cheng Kung University, Tainan, Taiwan.

[4]Princeton Institute for Science and Technology of Materials (PRISM), Princeton University, Princeton, New Jersey, USA.

[5]Department of Physics, Boston College, Chestnut Hill, Massachusetts, USA.

[6]Princeton Center for Theoretical Science, Princeton University, Princeton, New Jersey 08544, USA.

[7]Department of Physics, Princeton University, Princeton, New Jersey, USA.

[8]Institute of Physics, Academia Sinica, Taipei, Taiwan.

[9]Department of Physics, University of Zurich, Winterthurerstrasse, Zurich, Switzerland.

[10]Center for Quantum Frontiers of Research & Technology (QFort), Tainan, Taiwan.

[11]Physics Division, National Center for Theoretical Sciences, Hsinchu, Taiwan.

[12]CAS Center for Excellence in Topological Quantum Computation, University of Chinese Academy of Sciences, Beijing, China.

[13]Beijing Academy of Quantum Information Sciences, West Building 3, No. 10 Xibeiwang East Road, Haidian District, Beijing, China.

[14]Lawrence Berkeley National Laboratory, Berkeley, California 94720, USA.

†Corresponding authors, E-mail:

jiaxiny@princeton.edu; gwljiashuang@pku.edu.cn; mzhasan@princeton.edu;
*These authors contributed equally to this work.

**The quantum level interplay between geometry, topology, and correlation is at the forefront of fundamental physics[1-15]. Owing to the unusual lattice geometry and breaking of time-reversal symmetry, kagome magnets are predicted to support intrinsic Chern quantum phases[14,15]. However, quantum materials hosting ideal spin-orbit coupled kagome lattices with strong out-of-plane magnetization have been lacking[16-21]. Here we use scanning tunneling microscopy to discover a new topological kagome magnet $TbMn_6Sn_6$, which is close to satisfying the above criteria. We visualize its effectively defect-free purely Mn-based ferromagnetic kagome lattice with atomic resolution. Remarkably, its electronic state exhibits distinct Landau quantization upon the application of a magnetic field, and the quantized Landau fan structure features spin-polarized Dirac dispersion with a large Chern gap. We further demonstrate the bulk-boundary correspondence between the Chern gap and topological edge state, as well as the Berry curvature field correspondence of Chern gapped Dirac fermions. Our results point to the realization of a quantum-limit Chern phase in $TbMn_6Sn_6$, opening up an avenue for discovering topological quantum phenomena in the $RMn_6Sn_6$ (R = rare earth element) family with a variety of magnetic structures. Our visualization of the magnetic bulk-boundary-Berry correspondence covering real and momentum space demonstrates a proof-of-principle method revealing topological magnets.**

Exploring quantum topology under nontrivial lattice geometry and strong electron interaction is emerging as the new frontier in condensed matter physics, which not only has analogies to high-energy physics but also expands quantum materials for next-generation technology[1-15]. Recently, the transition metal-based kagome magnets have attracted great attention, as they often exhibit correlated topological band structures[7,8,16-20]. A kagome lattice, made of corner-sharing triangles, naturally possesses relativistic band crossings at the Brillouin zone corners (Fig. 1**a**). The inclusion of spin-orbit coupling and out-of-plane ferromagnetic ordering in the kagome lattice effectively realizes the spinless Haldane model generating Chern gapped topological fermions[9,10,13-15] (Fig. 1**b**). However, direct experimental visualization of the phenomenon remains challenging, due to the existence of Sn atoms in the kagome layer, the close stacking of the kagome lattices, and the tendency to form in-plane magnetization for most kagome magnets[16-20]. Through intense research, we find that $TbMn_6Sn_6$ provides key material advancement. Unlike other members of the kagome magnet family, it consists of segregated kagome layers formed purely by Mn atoms. More crucially, its kagome lattice uniquely features both an out-of-plane magnetization ground state and the largest coercivity (1.1T) within the $RMn_6Sn_6$ family (Fig. 1**c**, Ref. 22-24). Therefore, $TbMn_6Sn_6$ is a tantalizing system to search for the Chern gapped topological fermions.

$TbMn_6Sn_6$ has a layered crystal structure with space group P6/mmm and hexagonal lattice constants $a$ = 5.5 Å and $c$ = 9.0 Å. It consists of a Mn kagome layer with Sn and Tb successively distributed in alternating layers stacked along the $c$-axis. The material has a ferrimagnetic ground state (Curie temperature, $T_C$ = 423K), with a Mn moment of $2.4\mu_B$ ferromagnetically aligned along the $c$-axis and a Tb moment of $8.6\mu_B$ anti-aligned (Fig. 1**d**, ref. 24). We grow high-quality single crystals through a flux method. The crystal exhibits a well-defined magnetization loop for the field applied along the $c$-axis and no magnetization loop for the field applied within the $a$-$b$ plane in Fig. 1**d** (right-side), confirming the strong out-of-plane magnetization. The side-plane map of the crystal measured by scanning transmission electron microscopy in Fig. 1**e** directly demonstrates its atomic stacking sequence along the $c$-axis. It can be seen that the interlayer distance between the TbSn layer and Mn layer is largest, and the crystal tends to cleave along this plane. Searching extensively over 50 cryogenically cleaved crystals with low temperature (T = 4.2K) scanning tunneling microscopy, we were able to obtain the large atomically flat Mn kagome lattice, as shown in Fig. 1**f**. Its zoomed-in image, measured with a much lower junction resistance set up, directly reveals the Mn kagome atoms. Moreover, unlike topographies of the kagome lattice in other kagome magnets[7,8,20,25,26], which show various atomic defects, there is no detectable defect over a large field of view. The experimental visualization of such a defect-free magnetic kagome lattice offers an unprecedented opportunity to explore its intrinsic topological quantum properties.

Next, we measure the low-energy tunneling spectrum of the Mn kagome lattice under an applied magnetic field (Figs. 2**a** and **b**). We find the zero-field spectra to be spatially homogeneous. When applying a 9T magnetic field along the c-axis, the spectra show drastic changes with the emergence of a series of states widely distributed in energy, which is a clear signature of Landau quantization. We find the Landau quantization of the kagome layer is unique within this material; as for the other observed lattice with stripe morphology, we do not detect a strong field response up to 9T (Fig. 2**c** and **d**). This surface is likely to be the TbSn layer based on the aforementioned easy cleavage plane and the fact that $R^{3+}$ surface ions may have dangling bonds favoring reconstruction[27]. Moreover, none of the reported tunneling studies on kagome materials has shown Landau quantization[7,8,20,25,26]. Therefore, the Landau quantization of the magnetic Mn kagome lattice suggests that it is distinguishably in the quantum-limit.

To understand the origin of this Landau quantization, we map its fan diagram in Fig. 3**a** by slowly increasing the magnetic field. Mapping out the Landau fan is a nontrivial task in tunneling experiments, and there are only a few successful examples in quantum materials including graphene[28], bismuth[29], and topological insulators[30,31]. In these cases, analysis of the Landau fan extracts precise band structure information in momentum space, but applying such methodology to a correlated topological magnet remains challenging. For a spin-orbit coupled kagome lattice with out-of-plane magnetization, it is natural to consider the existence of spin-polarized Dirac fermions with a Chern gap[14,15]. We highlight several key features in the Landau fan diagram that constrain the analysis along this direction. First, the zero-field peak shifts almost linearly to lower energy with increasing field, which indicates the presence of magnetic polarization with a Zeeman term ($\Delta E = \frac{1}{2} g \mu_B B$, $\mu_B$ is the Bohr magneton). The observation of a Zeeman shift rather than a splitting demonstrates that the electronic states are spin-polarized, which is crucial for the Chern gap formation[14,15]. Secondly, below this state, the other Landau levels shift nonlinearly with a square root-like field dependence, and their separation at 9T decreases for levels at lower energies, both factors of which are consistent with Dirac like fermions[28] featuring the energy spectrum $\varepsilon_n \sim \sqrt{|n|B}$ ($n = 0, \pm 1, \pm 2 \ldots$). Thirdly, above the zero-field peak, an intense state emerges and shifts in parallel with it. These two states that shift linearly with field are likely to define the expected Chern gap $\Delta$ that is determined by the intrinsic spin-orbit coupling. The zero-field peak may be formed by accumulated states from the top of the lower Dirac branch[7].

A Chern gap modifies the bare Dirac dispersion $\varepsilon_k = E_D \pm \hbar k v$ into $E_k = E_D \pm \sqrt{(\Delta/2)^2 + (\hbar k v)^2}$ ($E_D$ is the Dirac cone energy, $\hbar$ is the reduced Planck's constant, $v$ is the Dirac velocity). Hence to describe this Landau fan diagram, we start with a formula written as: $E_n = E_D \pm \sqrt{(\Delta/2)^2 + 2|n|e\hbar v^2 B} - \frac{1}{2} g \mu_B B$. A simulation of the Landau fan data with this formula is shown in Fig. 3**b** with parameters: $E_D$ = 130 ± 4 meV, $\Delta$ = 34 ± 2 meV, $v$ = 4.2 ± 0.3×10⁵ m/s and $g$ = 52 ± 2. The large g factor has also been reported in other topological materials, which may arise from the orbital contribution. In the kagome tight-binding model with nearest-neighbor hopping, the Dirac dispersions appear at the Brillouin zone corners. Our exploration of the band structure below the Fermi level via angle-resolved photoemission indeed finds linear dispersions near the zone corners with the similar Fermi velocity, which also reasonably connect to the Chern gapped Dirac band extracted from the tunneling data, as shown in Fig. 3**c**. We further find that the gap extracted directly from the energy distance of the two peaks increases slightly with the field, correlating with that of the out-of-plane magnetization value $M_C$ (Fig. 3**d**). The weak field dependence supports the interpretation that the Chern gap is not opened by the external field but induced by the intrinsic spin-orbit coupling[15]. The existence of predominate Chern gapped Dirac fermions just around the Fermi level is another key factor in driving this defect-free kagome lattice to the quantum-limit.

The nontrivial topology of the Chern gap produces the dissipationless edge state. To visualize this bulk-boundary correspondence, we perform tunneling measurements to map a step edge in Fig. 4**a**. Both the upper and lower layers of the step edge are surfaces formed by the Mn kagome lattices with a unit cell step height of ~9Å, and therefore have similar density of states. We observe a pronounced localized edge when mapping at energy within the Chern gap, while no clear edge state is detectable at other energies outside the gap, confirming the existence of the nontrivial in-gap edge state[32,33]. We also explore the tunneling signal on the side cleaving surface (Fig. 4**b**). We perform mapping over a large area, and their direct Fourier transforms give rise to the quasi-particle scattering signal. We observe that quasi-particle scattering along the bulk edge direction is substantially reduced within the energy range of the Chern gap, in agreement with the dissipationless nature (lack of backscattering) of the Chern edge state. The magnetic Landau fan exhibiting a Chern gap and emergence of in-gap edge states lack of backscattering together provide spectroscopic evidence for the topological bulk-boundary correspondence.

In addition to the bulk-boundary correspondence, Chern gapped Dirac fermions will also feature large Berry curvature[34-40]. This Berry curvature contribution to anomalous Hall conductivity is estimated[39] as $\sigma_{xy} = \frac{\Delta}{2E_D} * e^2/h = 0.13 \pm 0.01 e^2/h$ based on the tunneling data. On the other hand, we observe the anomalous Hall signal $\rho_{AH}$ in the Hall resistivity of the bulk crystal (Extended Data Fig. 3). When plotting $\rho_{AH}$ against the square of the longitudinal resistivity $\rho_{xx}^2$, we observe a linear scaling (Fig. 4**c**), indicating a predominant intrinsic contribution[34-38]. From the linear fit ($\rho_{AH} = A + \sigma^{int}\rho_{xx}^2$), we find that this intrinsic contribution is $\sigma^{int}$ = 121 ± 6 Ω⁻¹cm⁻¹. It amounts to $\sigma_{xy}^{int} = 0.14 \pm 0.01\ e^2/h$ per Mn kagome layer, which agrees with expected $\sigma_{xy}$, attesting to the Berry curvature correspondence of Chern gapped Dirac fermions.

Our observations of magnetic Landau quantization and bulk-boundary-Berry correspondence provide strong evidence in space and momentum for a quantum-limit Chern magnet. It is extremely rare to find a topological magnetic system featuring the quantized Landau fan, which requires defect-free magnetic material design and cutting-edge spectroscopy characterization. It is equally rare to find a large Chern gap system to demonstrate its topological correspondence, which is one of the key pursuits in the pertinent fundamental research area[9,10]. Given that there are dozens of compounds with similar structures to $TbMn_6Sn_6$ that host kagome lattices with a variety of magnetic structures and wide tunability of the lattice constant, our findings can be a valuable guideline in discovering other intimately related, yet hitherto unknown topological or quantum phenomena.

**References:**

1. Keimer, B. & Moore, J. E. The physics of quantum materials. *Nature Physics* **13**, 1045–1055 (2017).
2. Sachdev, S. Topological order, emergent gauge fields, and Fermi surface reconstruction. *Rep. Prog. Phys.* **82**, 014001 (2019).
3. Franz, M. & Rozali, M. Mimicking black hole event horizons in atomic and solid-state systems. *Nature Reviews Materials* **3**, 491–501 (2018).
4. Tokura, Y., Yasuda, K. & Tsukazaki, A. Magnetic topological insulators. *Nature Review Physics* **1**, 126-143 (2019).
5. Hasan, M. Z. et al. Topological insulators, topological superconductors and Weyl fermion semimetals: discoveries, perspectives and outlooks. *Phys. Scr. T* **164**, 014001 (2015).
6. Armitage, N. P., Mele, E. J. & Vishwanath, A. Weyl and Dirac semimetals in three-dimensional solids. *Rev. Mod. Phys.* **90**, 015001 (2018).
7. Yin, J. X. *et al.* Giant and anisotropic many-body spin–orbit tunability in a strongly correlated kagome magnet. *Nature* **562**, 91–95 (2018).
8. Yin, J.-X. *et al.* Negative flat band magnetism in a spin–orbit-coupled correlated kagome magnet. *Nature Physics* **15**, 443–448 (2019).
9. Thouless, D. J. *et al.* Quantized Hall Conductance in a Two-Dimensional Periodic Potential. *Phys. Rev. Lett.* **49**, 405 (1982).
10. Haldane, F. D. M. Model for a Quantum Hall Effect without Landau Levels: Condensed-Matter Realization of the "Parity Anomaly". *Phys. Rev. Lett.* **61**, 2015–2018 (1988).
11. Chang, C.Z. *et al.* Experimental Observation of the Quantum Anomalous Hall Effect in a Magnetic Topological Insulator. *Science* **340**, 167-170 (2013).
12. Sharpe, A. L. *et al.* Emergent ferromagnetism near three-quarters filling in twisted bilayer graphene. *Science* **365**, 605-608 (2019).
13. Zou, J., He, Z. & Xu, G. The study of magnetic topological semimetals by first principles calculations. *npj Comput Mater* **5**, 96 (2019).
14. Tang, E., Mei, J. W. & Wen, X. G. High-temperature fractional quantum Hall states. *Phys. Rev. Lett.* **106**, 236802 (2011).
15. Xu, G., Lian, B. & Zhang, S.-C. Intrinsic Quantum Anomalous Hall Effect in the Kagome Lattice $Cs_2LiMn_3F_{12}$. *Phys. Rev. Lett.* **115**, 186802 (2015).

16. Nakatsuji, S., Kiyohara, N. & Higo, T. Large anomalous Hall effect in a non-collinear antiferromagnet at room temperature. *Nature* **527**, 212–215 (2015).
17. Ye, L. *et al.* Massive Dirac fermions in a ferromagnetic kagome metal. *Nature* **555**, 638–642 (2018).
18. Liu, E. *et al.* Giant anomalous Hall effect in a ferromagnetic kagome-lattice semimetal. *Nature Physics* **14**, 1125–1131 (2018).
19. Kang, M. *et al.* Dirac fermions and flat bands in the ideal kagome metal FeSn. *Nature Materials* **19**, 163–169 (2020).
20. Lin, Z. *et al.* Flatbands and emergent ferromagnetic ordering in $Fe_3Sn_2$ kagome lattices. *Phys. Rev. Lett.* **121**, 096401 (2018).
21. Kane, C. L. & Mele, E. J. Quantum spin Hall effect in graphene. *Phys. Rev. Lett.* **95**, 226801 (2005).
22. Venturini, G., ElIdrissi, B. C. & Malaman, B. Magnetic properties of $RMn_6Sn_6$ (R = Sc, Y, Gd−Tm, Lu) compounds with $HfFe_6Ge_6$ type structure. *Journal of Magnetism and Magnetic Materials* **94**, 35-42 (1991).
23. Malaman, B. *et al.* Magnetic properties of $RMn_6Sn_6$ (R = Gd–Er) compounds from neutron diffraction and Mössbauer measurements. *Journal of Magnetism and Magnetic Materials* **202**, 519-534 (1999).
24. ElIdrissi, B. C., Venturini, G. & alaman, B. Magnetic structures of $TbMn_6Sn_6$ and $HoMn_6Sn_6$ compounds from neutron diffraction study. *Journal of the Less-Common Metals* **175**, 143-154 (1991).
25. Jiao, L. *et al.* Signatures for half-metallicity and nontrivial surface states in the kagome lattice Weyl semimetal $Co_3Sn_2S_2$. *Phys. Rev. B* **99**, 245158 (2019).
26. Yang, H.-H. *et al.* Scanning tunneling microscopy on cleaved $Mn_3Sn(0001)$ surface. *Sci. Rep.* **9**, 9677 (2019).
27. Rößler, S. *et al.* Hybridization gap and Fano resonance in $SmB_6$. *Proc. Natl Acad. Sci.* **111**, 4798–4802 (2014).
28. Miller, D. L. *et al.* Observing the quantization of zero mass carriers in graphene. *Science* **324**, 924-927 (2009).
29. Feldman, B. E. *et al.* Observation of a Nematic Quantum Hall Liquid on the Surface of Bismuth. *Science* **354**, 316-321 (2016).
30. Okada, Y., Serbyn, M., Lin, H. & Walkup, D. Observation of Dirac node formation and mass acquisition in a topological crystalline insulator. *Science* **341**, 1496–1499 (2013).
31. Hanaguri, T., Igarashi, K. & Kawamura, M. Momentum-resolved Landau-level spectroscopy of Dirac surface state in $Bi_2Se_3$. *Phys. Rev. B* **82**, 081305 (2010).
32. Wu, R. *et al.* Evidence for Topological Edge States in a Large Energy Gap near the Step Edges on the Surface of $ZrTe_5$. *Phys. Rev. X* **6**, 021017 (2016).
33. Tang, S. *et al.* Quantum spin Hall state in monolayer 1T'-WTe2. *Nature Physics* **13**, 683–687 (2017).
34. Tian, Y. *et al.* Proper scaling of the anomalous Hall effect. *Phys. Rev. Lett.* **103**, 087206 (2009).
35. Nagaosa, N., Sinova, J., Onoda, S., MacDonald, A. H. & Ong, N. P. Anomalous Hall effect. *Rev. Mod. Phys.* **82**, 1539-1592 (2010).
36. Haldane, F. D. M. Berry Curvature on the Fermi Surface: Anomalous Hall Effect as a Topological Fermi-Liquid Property. *Phys. Rev. Lett.* **93**, 206602 (2004).

37. Yates, J. *et al.* Spectral and Fermi surface properties from Wannier interpolation. *Phys. Rev. B* **75**, 195121 (2007).
38. Chang, M-C. *et al.* Berry phase, hyperorbits, and the Hofstadter spectrum: Semiclassical dynamics in magnetic Bloch bands. *Phys. Rev. B* **53**, 7010 (1996).
39. Sinitsyn, N. A., MacDonald, A. H., Jungwirth, T., Dugaev, V. K. & Sinova, J. Anomalous Hall effect in a two-dimensional Dirac band: The link between the Kubo-Streda formula and the semiclassical Boltzmann equation approach. *Phys. Rev. B* **75**, 045315 (2007).
40. Kou, X. *et al.* Scale-invariant quantum anomalous Hall effect in magnetic topological insulators beyond the two-dimensional limit. *Phys. Rev. Lett.* **113**, 137201 (2014).

**Acknowledgements** We thank Brian Sales and Yimin Xiong for providing other kagome materials for comparison study. We thank P. W. Anderson, David Huse, Duncan Haldane, Sanfeng Wu, and Nai Phuan Ong for insightful discussions. Experimental and theoretical work at Princeton University was supported by the Gordon and Betty Moore Foundation (GBMF4547/ Hasan). The material characterization (ARPES) is supported by the United States Department of energy (US DOE) under the Basic Energy Sciences programme (grant number DOE/BES DE-FG-02-05ER46200). The work in Peking University was supported by the National Natural Science Foundation of China No. U1832214, No.11774007, the National Key R&D Program of China (2018YFA0305601) and the strategic Priority Research Program of Chinese Academy of Sciences (XDB28000000). Work at Princeton's Imaging and Analysis Center is supported by the Princeton Center for Complex Materials, a National Science Foundation (NSF)-MRSEC program (DMR-1420541). Work at Boston College was supported by the U.S. Department of Energy, Basic Energy Sciences Grant No. DE-FG02-99ER45747. T.N. acknowledges supports from the European Union's Horizon 2020 research and innovation program (ERC-StG-Neupert-757867-PARATOP). T.A.C. acknowledges support by the National Science Foundation Graduate Research Fellowship Program under Grant No. DGE-1656466. This research used resources of the Advanced Light Source, a DOE Office of Science User Facility under contract No. DE-AC02-05CH11231. The authors thank Sung-Kwan Mo for support at beamline 10.0.1 of the Advanced Light Source. We acknowledge Diamond Light Source for time on beamline i05 under Proposal SI22332-1. The authors thank Cephise Cacho and Timur Kim for support at beamline i05 of Diamond Light Source. T.-R.C. was supported by the Young Scholar Fellowship Program from the Ministry of Science and Technology (MOST) in Taiwan, under a MOST grant for the Columbus Program MOST109-2636-M-006-002, National Cheng Kung University, Taiwan, and National Center for Theoretical Sciences, Taiwan. This work was supported partially by the MOST, Taiwan, Grant MOST107-2627-E-006-001. This research was supported in part by Higher Education Sprout Project, Ministry of Education to the Headquarters of University Advancement at National Cheng Kung University (NCKU). M.Z.H. acknowledges support from Lawrence Berkeley National Laboratory and the Miller Institute of Basic Research in Science at the University of California, Berkeley in the form of a Visiting Miller Professorship.

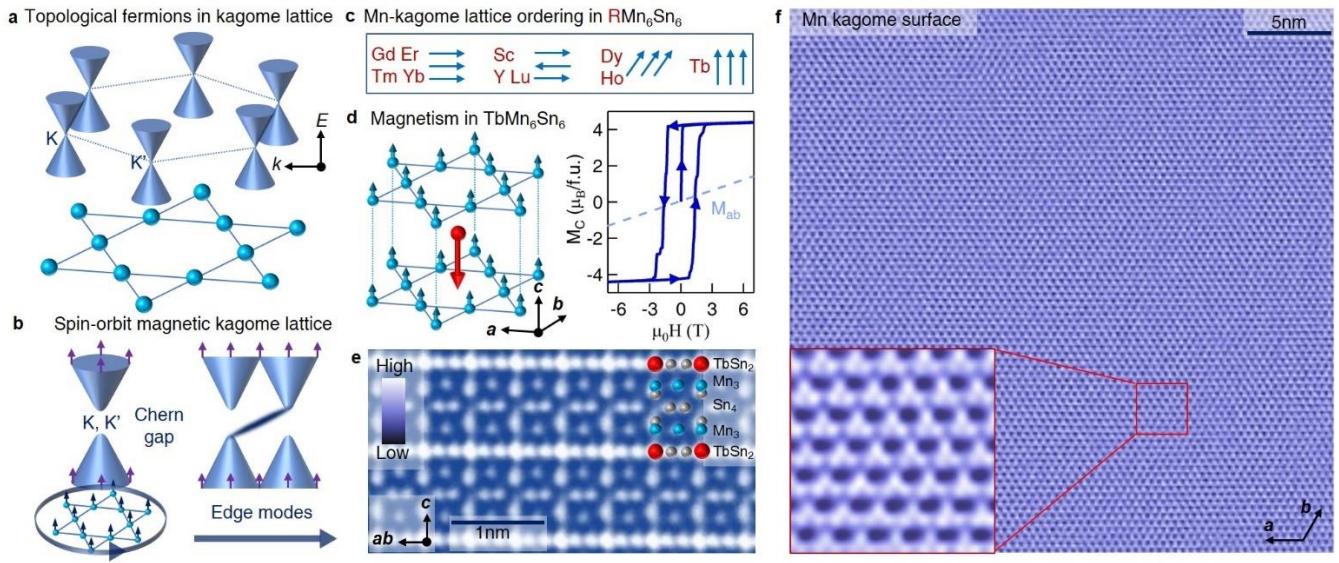

**Fig. 1 Atomic-scale visualization of the defect-free magnetic kagome lattice.** a, Illustration of Dirac band crossings (hourglass cones) at the Brillouin zone (dash-lines) corners for a kagome lattice (spheres connected by solid lines). b, Left: illustration of spin-polarized Dirac fermions with a Chern gap (two separated cones with arrows) in the spin-orbit coupled magnetic kagome lattice (spheres with arrows). Right: illustration of the edge mode (purple line) that arises within the Chern gap. c, Summary of the magnetic ground state of the Mn kagome lattice in the $RMn_6Sn_6$ family, including in-plane ferromagnetism (R = Gd, Er, Tm, Yb), in-plane antiferromagnetism (R = Sc, Y, Lu), canted ferromagnetism (R = Dy, Ho), and out-of-plane ferromagnetism (R = Tb). d, Magnetism in $TbMn_6Sn_6$ with the left image illustrating its magnetic structure of Mn (blue) and Tb (red) atoms, and the right image showing the out-of-plane (solid-line) and in-plane (dash-line) magnetization curves taken at 4.2K. e, Scanning transmission electron microscope image of $TbMn_6Sn_6$, showing the atomic interlayer stacking. f, Scanning tunneling microscopy image of the Mn terminating surface taken at 4.2K. The inset shows the magnified image of the kagome lattice.

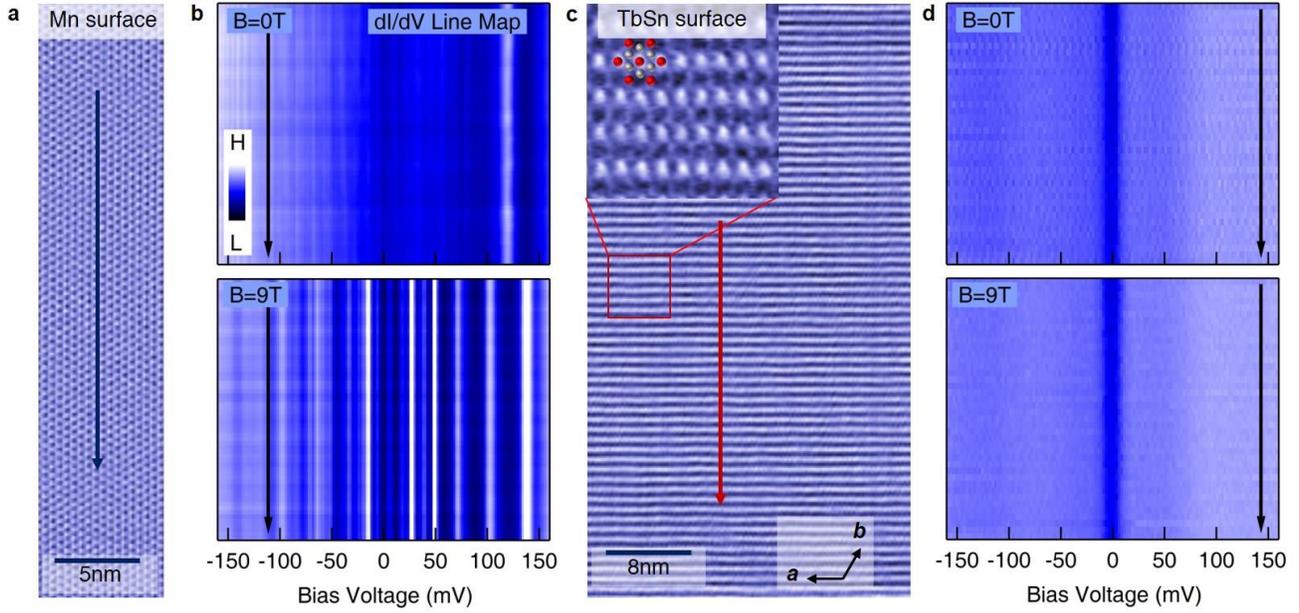

**Fig. 2 Distinct Landau quantization of Mn kagome lattice.** a, Topography of the Mn kagome lattice. b, Corresponding dI/dV line maps (along the line in a) taken at B=0T (upper panel) and B=9T (lower panel), respectively. The 9T data shows intense modulation, which is associated with Landau quantization. c, Topography of the striped surface. Inset: magnified atomic view overlaid with schematic TbSn lattice. d, Corresponding dI/dV line maps (along the red line in c) taken at B=0T (upper panel) and B=9T (lower panel), respectively.

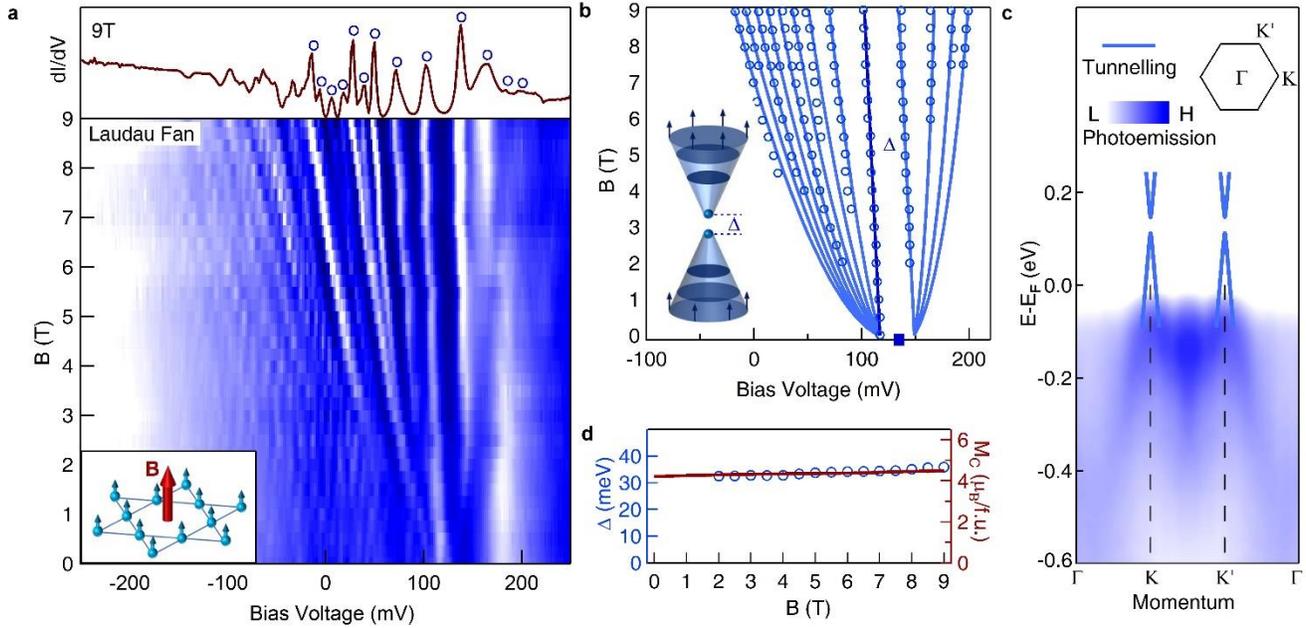

**Fig. 3 Quantum-limit visualization of Chern gapped Dirac fermions.** a, Landau fan diagram of the kagome lattice. The inset illustrates the magnetic field applied perpendicular to the kagome lattice. The upper panel shows the 9T dI/dV spectrum, with Landau levels marked by open circles. b, Fitting the Landau fan data (open circles) with the spin-polarized and Chern gapped Dirac dispersion (solid lines). Inset: schematic of Landau quantization of Chern gapped Dirac fermions. c, Comparison of the dispersion obtained from tunneling and photoemission. The inset shows the Brillouin zone for a kagome lattice. d, Dirac gap size (open circles) and out-of-plane magnetization (solid line) as a function of the magnetic field.

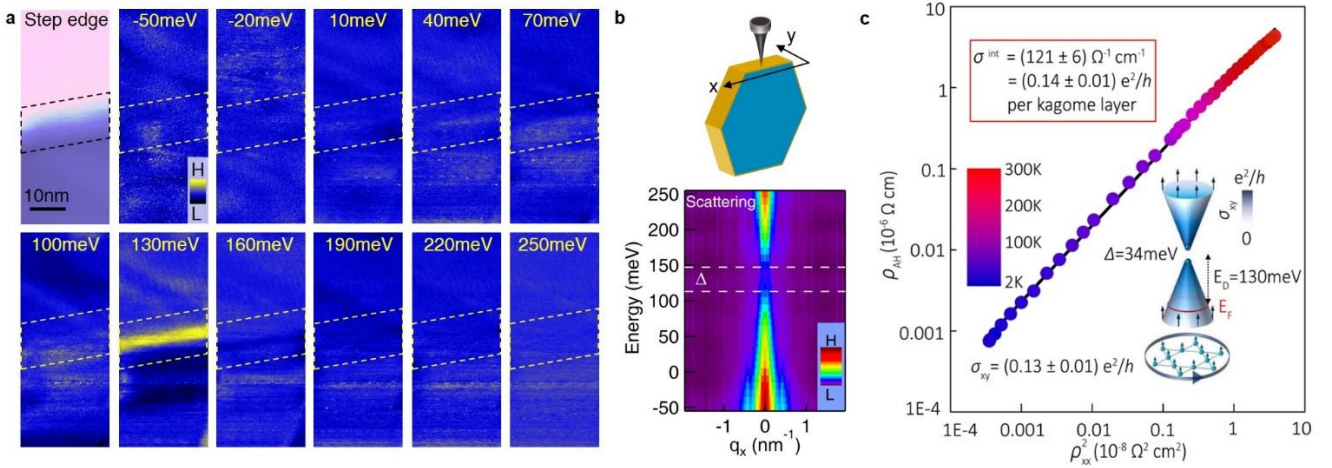

**Fig. 4 Correspondence of Chern gapped Dirac fermions with topological edge state and Berry curvature.** a, dI/dV maps taken at different energies across a step edge (left up panel). The map taken within the Chern gap energy (130meV) shows a pronounced step edge state. b, Quasi-particle scattering along the bulk crystal edge direction (illustrated on top). The white lines in the scattering map mark the Chern gap energy determined by the Landau quantization. c, The anomalous Hall resistivity $\rho_{AH}$ plotted against $\rho_{xx}^2$ in a log scale from 2K to 300K. The intrinsic Hall conductance is given by the slope of the line, which amounts to $0.14 \pm 0.01$ $e^2/h$ per Mn kagome layer (upper inset). The lower inset illustrates the Berry curvature contribution to the Hall conductivity from Chern gapped Dirac fermions, which is $0.13 \pm 0.01$ $e^2/h$ per Mn kagome layer based on the tunneling data.

10